\documentclass[a4paper,11pt]{article}
\def\ll{\label}
\def\re{\ref}
\def\c{\cite}

\def\r1{(\ref{$1})}

\def\sn{\rm sn}

\def\cn{\rm cn}
\def\dn{\rm dn}

\def\th{\theta}
\def\ba{\begin{array}{c}}

\def\ea{\end{array}}

\def\si{\sigma}

\def\bet{\beta}
\def\ov{\over}
\def\ha{{1\over 2}}

\def\l{\left}
\def\l({\left(}
\def\r){\right)}
\def\r{\right}

\def\la{\lambda}
\def\al{\alpha}

 \def\be{\begin{equation}}
\def\bc{\begin{center}}
\def\ec{\end{center}}
\def\bit{\begin{itemize}}
\def\eit{\end{itemize}}
\def\ee{\end{equation}}
\def\ed{\end{document}}
\def\bea{\begin{eqnarray}}
\def\eea{\end{eqnarray}}
\def\efr{\end{flushright}}

\begin{document}
\title{Quantum integrable multi atom matter-radiation models with and without
rotating wave approximation}

\author{
Anjan Kundu \\  
  Saha Institute of Nuclear Physics,  
 Theory Group \\
 1/AF Bidhan Nagar, Calcutta 700 064, India.
\\  {email: anjan@tnp.saha.ernet.in} 
\footnote{Talk presented at NEEDS04, Gallipoli, Italy, July 2004 }
}
\maketitle
\begin{abstract} 
New  integrable multi-atom  matter-radiation models with and without
rotating wave approximation (RWA) are constructed and exactly solved through
  algebraic Bethe ansatz. The models with RWA are generated through ancestor
model approach in an unified way.  The rational case yields the standard
type of matter-radiaton models, while the trigonometric case corresponds to
their q-deformations.  The models without RWA are obtained from the elliptic
case at the Gaudin and high spin limit.

 42.50 Pq,
03.65 Fd, 
32.80 -t 

\end{abstract}
\section {Introduction}
It is amazing to note that real
   systems, like
 those in quantum optics induced by
resonance interaction  between atom and a quantized  laser field, 
 in cavity QED
 both in microwave and optical domain 
\c{rempe8790,raizen89}, 
in trapped ion
interacting with its center of mass motion
 irradiated by a laser beam \c{trap,vogel95} etc.,
all involving      complex   matter 
radiation (MR) interactions  can be  described  so  successfully by 
 simple  models,
 like   Jaynes-Cummings (JC)  \c{jc} and
 Buck Sukumar (BS)  \c{bs} models.  
Many theoretical predictions  based on these models, like
 vacuum Rabi splitting 
, Rabi oscillation  and its  quantum collapse and revival
etc. 
 have been verified in maser and laser experiments.

In such simplest  interacting MR models the quantized radiation
 field is taken in  single bosonic
mode: 
$h=b e^{-i\omega_ft}+b^\dag e^{i\omega_ft}$, while  the atom is considered 
to be  a two-level
spin system  with polarization vector $S=\si^- e^{-i\omega_at}+\si^+
 e^{+i\omega_at}$. 
The interaction $(h\cdot S)$ therefore
 contains in general both fast
(with frequency $w_f+w_a$) and slow oscillating 
(with frequency $w_f-w_a$) components. However
it is 
customary to neglect the fast oscillating part 
by considering  approximation
 $|w_f-w_a| << w_f $, which corresponds to the  rotating wave
 approximation (RWA) and  make the model solvable.
This yields  with the addition of free field and atomic excitation terms 
the well known JC model 
\be H_{JC}= \omega_f b^\dag b+ { \omega_a} \si^z+ 
\al (b^\dag \si^-+ b \si^+).  
 \ll{rw}\ee
However this  approximation, which
  is justified only near the resonance point $ \omega_f \approx { \omega_a} $,
  should be avoided in the
general case \c{rwa}, when additional 
counter rotating wave (CRW) terms must appear: 
\be
H_{crw}= 
 \bet (b^\dag \si^++ b \si^-)  
 \ll{crw}\ee
  Moreover, for describing physical situations more
 accurately, one has to look
  for  generalizations of the basic models, i.e. has to  consider 
their q-deformations
 \c{qjc,qbs}, introduce higher nonlinearities
\c{vogel95,ntrap}, involve
 multiple atoms \c{raizen89,ntrap,natom} etc.
  However, while the exact solutions for the JC and
the BS models together with their simple multi-atom extensions
 are known \c{jcexact,jcbethe,bsbethe},  the same appears to be no longer true
for most of the above generalizations. More
 precisely, 
integrable  models without RWA as well as those with explicit inter-atomic
 couplings   are not known for most of the MR models,
 except for a few recent attempts \c{amic04,duk04}.
   Moreover, though q-deformation,  signifying introduction of anisotropy
 together with specific nonlinearity into the system, was considered earlier
for a few  models \c{qjc,qbs}, their multi-atom and integrable variants
  are not  known.
 Therefore, it is indeed a challenge to find a  scheme
for generating integrable  MR models having 
the desired properties mentioned above.

\section{Major aims and strategy of our construction}

 To meet this challenge we construct 
new classes of integrable MR models with and without RWA.
The result concerning our models with RWA has been
 reported recently \c{jpa04}, where we propose a general integrable system
 based on our ancestor Lax operator approach \c{kunprl}.  Through various
reductions we generate from it a series of integrable MR models with
explicit inter-atomic interactions.  This includes such new multi-atom
 generalizations of JC, BS and trapped ion (TI) models as well as their
 integrable q-deformations involving quantum group.
Moreover, since   the construction of our
  models is based on  a general Yang-Baxter (YB) algebra, we
  can also solve them in a unified way through the Bethe ansatz (BA).  Our
aim here is to review briefly this result  and then  
 present our  new result obtained without RWA.

Our strategy for constructing    models with RWA  is to start with 
  a  combination of Lax operators: 
 $T(\la)=L^{s}(\la) \prod_j^{N_a}L_j^{S}(\la)$,
where $L^{s}(\la)$ involving   bosons represents the  
 radiation or the oscillatory mode and 
$L_j^{S}(\la)$ involving spin operators represents  $N_a$-number of atoms.
 By construction  
it must satisfy the quantum YB  equation (QYBE)
 $R(\la-\mu)T(\la)\otimes T(\mu)= (I\otimes T(\mu))
(T(\la)\otimes I)R(\la-\mu)$, yielding the integrability condition 
 $[\tau (\la),\tau (\mu)]=0$, with mutually commuting set of conserved
operators, 
  obtained through  expansion   $
\tau (\la)= tr T(\la)=\sum_a C_a \lambda ^a $ \c{aba}.
For constructing the bosonic Lax operator  we implement  our ancestor model
approach \c{kunprl} and from a generalized Lax operator
 linked with a  generalized
   quadratic  algebra or its quantum deformation  generate different $L^{s}(\la)$ through  various bosonic realizations.  
Our  standard MR models 
 are   linked to  the Lie algebra and the   rational  
  $R$-matrix of the
$xxx$  spin chain \c{aba}, while their    $q$-deformations  
are   related to the quantum algebra and the 
 trigonometric  $R$-matrix of  the $xxz$ chain  \c{xxz}.

 For  constructing    integrable MR  models without RWA, however, as we
 explain below,
   the above approach fails due to the difficulty in bosonizing 
 the
 Sklyanin algebra . Therefore for such models, 
which are linked to the elliptic
 $R$-matrix  of   the $xyz$ spin chain,  we have to change our strategy
and switch over  to the Gaudin limit together
 with a high spin limit.

\section {Integrable MR models with RWA}
We concentrate first on standard MR models and recall that in the rational
case the $2\times 2$ ancestor Lax operator may be given as \c{kunprl}
\be
L^{s}{(\la)} = \left( \begin{array}{c}
 {c_1^0} (\la + {s^3})+ {c_1^1}, \ \ \quad 
  s^-  \\
    \quad  
s^+ ,  \quad \ \ 
c_2^0 (\la - {s^3})- {c_2^1}
          \end{array}   \right), \ll{LK} \ee
with operators $ {\bf s}$ satisfying  a 
 quadratic algebra 
\be  [ s^+ , s^-] 
=  2m^+ s^3 +m^-,\ 
  ~ [s^3, s^\pm]  = \pm  s^\pm,\ \ [m^\pm,\cdot]=0. \ll{ralg}\ee 
The central elements   $m^\pm $ are 
expressed through
 arbitrary parameters appearing
 in (\re{LK}) as $m^+=c_1^0c_2^0,\ \  m^-= c_1^1c_2^0+c_1^0c_2^1$ and as  
 it is easy to see, their different  choices  reduce (\re{ralg})
 to  different simple algebras:
\bea
\mbox {i)  } su(u),  \ \mbox {at }  \ 
  m^+=1, m^-=0, \quad  
 \mbox {ii)  } su(1,1),  \ \mbox {at }  \ 
  m^+=-1, m^-=0,  \nonumber \\  
 \mbox {iii)  bosonic}, \ \mbox {at }  
  m^+=0, m^-=-1, \ \mbox {iv)  canonical},  \mbox {at }  
  m^+= m^-=0  \ll{algs}\eea 
and the corresponding limits yield  from (\re{LK})  the  respective Lax
operators. In case i), choosing $c_a^0=1$ and $ {c_1^1}=- {c_2^1}=
 {c_j} $, (\re{LK}) we get  the spin Lax operator
 $L^{S}_j{(\la)}, j \in [1,N_a] $ describing $N_a $ atoms with
inhomogeneity parameters $
 {c_j} $ , while
 the other  cases can reduce to different types of
 bosonic Lax operators linked to the radiation mode.

Remarkably, 
such reductions yield in a unified way
new integrable multi-atom BS, JC and  TI models,
  at the  limits ii), iii) and iv) of (\re{algs}). 
 Thus  case ii) with  choice   
$ c_1^0=- c_2^0=1,  c_1^1=c_2^1 \equiv c$,
  yields 
\be
H_{bs}= \omega_f s^3+ \sum_j^{N_a} \left( { \omega_a}_j S^z_j+
 \al (s^+ S^-_j+ s^- S_j^+)\right)  
+   \al \sum_{i < j}^{N_a} (  S^-_iS^+_j- S^+_iS^-_j),
 \ll{nbsh}\ee
 which with 
a bosonic 
realization of $su(1,1)$:  $ \ 
 s^+=\sqrt N b^\dag, 
s^-= b \sqrt N , s^3=N+\ha \ $ and taking the spin-$s$ operator as  
${\vec S}=\ha \sum_k^{2s}{\vec \sigma}_k$,
 represents a new {\it integrable multi-atom  
 BS model} with inter-atomic interactions and nondegenerate 
atomic frequencies. 
At $N_a=1$ the 
 matter-matter interactions obviously vanish,
  recovering
   the known    model  \c{bsbethe}.
The radiation frequency $\omega_f $ and the atomic frequencies 
$\omega_{aj} $ in our models are defined in general through the
inhomogeneity parameters  as 
\be \omega_f=\sum_j w_j, \ \ w_j=\al (c^0_1-c^0_2) c_{j}
, \ \ \  \omega_{aj}= \omega_f-w_j +\al (c^1_1+c^1_2)\ll{frec}\ee

Similarly,  under reduction
iii), choosing $c_1^0=\al, c_2^0=0, c_1^1 \equiv 
c, c_2^ 1=-\al^{-1}  $ and  taking direct bosonic
realization 
$s^-=b, s^+=b^\dag, s^3= b^\dag b ,$
  we obtain  from   the same (\re{LK}) an
  {\it  integrable multi-atom JC model}
 with matter-matter coupling.

From reduction iv),
by fixing the parameter values as $c_1^0=-1, 
 c_1^1 \equiv  c, c_2^ 0= c_2^1=0  $ and considering a
  consistent realization 
through canonical variables: $s^\pm=e^{\mp i x}, \ \ s^3=p+ x$,
we  generate an {\it integrable multi-atom TI model}
with full exponential nonlinearity:  $\al (e^{-ix} S^++e^{ix }S^-) $.
Such models however are difficult to solve 
by standard BA, due to the absence  of  pseudovaccum like in the Toda chain.

For constructing     
 {\it integrable q-deformed MR models} the strategy is the same; only
  one has to start  now
from the  trigonometric type ancestor Lax operator 
involving  generalized q-deformed  operators \c{kunprl}.
This yields an integrable system with Hamiltonian in the form
\bea
H_{qMR}&=& H_d+  (s_q^+ S^-_q+ s^-_q S_q^+)\sin \al ,  \nonumber \\ 
H_d&=&-ic_0 \cos (\al X) +c \sin (\al X), \  X=(s_q^3-S_q^z+\omega),
 \ll{qimrh}\eea
which represent  a new class of {  MR models} with ${\bf S}_q$ belonging to
the   quantum group
$U_q(su(2))$ and   ${\bf s}_q$ to a more general quantum  algebra
\c{kunprl}.

 {\it Integrable  q-deformed BS model}
is obtained from (\re{qimrh})       
 by realizing ${\bf s}_q$ through q-oscillator:
$s_q^+=\sqrt {[N]_q} b^\dag_q, \ s_q^-=b_q\sqrt{ [N]_q}, \ s_q^3=N+\ha  $,
and  using quantum spin operator  ${\bf S}_q$ 
 as its  co-product  \c{xxz} : 
$S^\pm_q=\sum_j^{s} q^{-\sum_{k<j}\sigma^z_k}
\sigma^\pm_jq^{\sum_{l>j}\sigma^z_l}, \ S^z =\sum_j^{s}\sigma^z_j$.   
At  $s=1$,  one gets an integrable
  version of an earlier  model \c{qbs}.

Similarly   realizing 
 $s_q^+= b^\dag_q, s_q^-=b_q, s^3=N $
yield a new {\it integrable q-deformation of the JC model}, and
 realization through canonical operators 
 results an {\it integrable q-deformed  TI
model}.

 We emphasize again that  all 
integrable MR models we proposed,
 similar to their unified construction,  allow
  exact  BA   solutions also  in a unified
 way.
More details on the above  models and their exact solutions can be found
  in \c{jpa04}.

\section {Integrable MR models without RWA}
 The basic idea for constructing  integrable matter-radiation 
models with RWA, as we have explained above,  is to   consider 
 first the $xxx$ or the $xxz$  spin chain with arbitrary spins and then
replace
one of the spins
 by its bosonic realization. The first case 
 belongs to the rational class, where the   
 spins satisfy the Lie type algebra, while  the second case falls in  the
  trigonometric class, where  the higher  spin operators
 satisfy the quantum algebra.
Fortunately both these algebras allow bosonic realization, which yields
the required radiation mode.
  
 For constructing more general 
integrable models with additional CRW terms (like
(\re{crw}) and its multiatom generalizations), one may 
 expect  to
apply  the same idea, but for  the $xyz$ spin chain.
 However in this
anisotropic  case,  linked to  the elliptic $R$-matrix, the higher spin
operators satisfy the Sklyanin algebra \c{sklyalg}, for which
unfortunately no bosonic realization is known.  In search for a way
out, one observes that
 the Sklyanin algebra fortunately 
 reduces again to the usual $su(2)$ algebra 
 at the elliptic Gaudin limit \c{egaudin},  allowing the required 
bosonization.
However, a direct bosonic realization of a single   spin operator in 
 Gaudin models  fails again , since it 
results either  nonintegrable  or unphysical models. The situation is saved
finally by a   further 
limit of $s \to \infty$, for the bosonic mode, 
which   yields the desired   integrable multi-atom JC type model with
nontrivial CRW
terms. 
We should mention that similar high spin limit has been used recently for
constructing integrable multi-atom JC model, but with RWA \c{duk04}.
  \subsection{Elliptic Gaudin model}
Since the first step in our construction is the elliptic Gaudin model (EGM), we
review briefly the related results from \c{egaudin}.
EGM is obtained at the $\alpha \to 0$ limit from the inhomogeneous 
$xyz$ spin model and therefore all  relevant objects like the Lax operator,
$R$-matrix, 
conserved quantities, as well as    
important  formulas in the Bethe ansatz
method, like the eigenfunctions and the eigenvalues  together with the Bethe
equations etc. concerning this model can be derived directly from those of the 
$xyz$ model \c{xyz} at the said limit.  The  Lax operator  with 
inhomogeneity parametrs $z_n$ and the 
$r$-matrix of the EGM can be expressed through 
 elliptic functions
in    the form
 \c{egaudin}
\bea
{L}_{n}(\la)&=& \sum_{a=1}^3 w_a (\la-z_n)  S^a_n \otimes \si^a , \ 
{r}(\la)= \sum_{a=1}^3 w_a (\la)  \si^a \otimes  \si^a , \ n=0, 1, \ldots, N_a
  \nonumber \\ 
\mbox{where} & & w_1(\la)={ \cn \ov \sn}(\la), \ 
 w_2(\la)={ \dn \ov \sn}(\la), \
 w_3(\la)={ 1 \ov \sn}(\la).
 \ll{Legaudin}\eea
 Operators
(\re {Legaudin})
satisfy a semiclassical YBE:
$[{L}_n(\la)\otimes{L}_n(\mu)] = [r(\la-\mu),
( {L}_n(\la)\otimes I+ I \otimes {L}_n(\mu)]$
,
 obtained at $\alpha \to 0$ limit from the QYBE.
Mutually commuting  conserved quantities $H_n$ of the EGM are obtained 
from the expansion coefficient of the $xyz $ transfer matrix 
\be 
\tau_{xyz}(\la,\al)=I+
 \al ^2 \tau^{(2)}(\la) +O(\al ^3)
\ll{tau}\ee
as 
\be\tau^{(2)}(\la \to z_n)=H_n=\sum_{a,m \neq n} w_a(z_n-z_m) S^a_nS^a_m
\ll{hegaudin}\ee
with $\sum_n H_n=0$.

It is most crucial for our purpose to note that,  higher spin operators 
$S^a_n$ appearing in the Gaudin model (\re{Legaudin}), are reduced 
from the generators  of the 
Sklyanin algebra to simply  those of 
\be  [ S^+_n , S^-_m] 
=  2\rho \delta_{nm}S^3_n ,\ 
  ~ [S^3_n, S^\pm_m]  = \pm \delta_{nm} S_n^\pm, \ll{su2}\ee 
with $\rho =\pm$ corresponding to $su(2)$ (or $su(1,1)$).
Recall that in constructing standard MR models with RWA we have taken
the spin operators for $N_a$ number of atoms as  generators of  $su(2)$
, while for the radiation field we have considered bosonic realizations
of  different algebraic structures  in (\re{algs}). For EMG  (\re{Legaudin})
however apart from $su(2)$, only other possible algebra is $su(1,1)$, as
given in (\re{su2}). And even this can not be used, since differnt algebras
can not be mixed in integrable Gaudin model. On the other hand, if we choose
$su(2)$ algebra for all operators and use the Holstein-Primakoff
transformation (HPT)
$
S^+_0= b^\dag \sqrt{ {s_0 \ov 2}-N }, \
S^-_0= \sqrt{ {s_0 \ov 2}- N} b  , \ S_0^3=N- {s_0 \ov 4}, \ N=b^\dag b
$ 
 for bosonizing  the
 radiative mode, though we can retain the integrability, but
 get unphysical Hamiltonian for $ <|N|> \  > {s_0 \ov 2}$,
 with $S^\pm_0$ becoming 
nonhermitian!
\subsection{High spin limit}
Fortunately, the above difficulties can be bypassed again by considering
further a high spin limit
 $s_0 \to \infty $  for the radiation field,
which reduces  the HPT to 
\be 
S^+_0={ {1 \ov  \sqrt 2\epsilon }} b^\dag, \
S^-_0=  { {1 \ov  \sqrt 2\epsilon }} b  , \ S^3_0=- {1 \ov  4 \epsilon^2}, \quad 
\epsilon = {1 \ov  \sqrt s_0} \to 0,
 \ll{hpt1}\ee
  by retaining terms up to order
$O({1\ov \epsilon})$.

For deriving now the  integrable matter-radiation  models without RWA  
 , we replace spin operators at
$n=0$ by bosons
 for the radiative mode  in (\re {hegaudin})  by using  (\re {hpt1})
and choose the arbitrary
parameters at $ n=0,j$ as \be
z_0=K+\epsilon x_0, \ z_j=\epsilon x_j, \ j=1,\ldots N_a  
,\ll{z2x}\ee
where $K$ is the elliptic integral \c{xyz} related to a  period of the elliptic 
functions. 
At the limit $\epsilon  \to 0 $ therefore  the coupling parameters 
$w_a(z_n-z_m) $  
can be expanded as
\bea
w_\pm(z_0-z_k=-w_\pm(z_k-z_0) \to  
\pm {k^{'}\ov 2} +O(\epsilon ), \ w_3(z_0-z_k)=-w_3(z_k-z_0) \to 
1 +O(\epsilon ^2), & & \nonumber \\
w_+(z_j-z_k) \to  
{1 \ov \epsilon} ({1\ov x_j-x_k}) +O(\epsilon ), \
w_-(z_j-z_k) \to  O(\epsilon ), \
 w_3(z_j-z_k) \to 
{1 \ov \epsilon} ({1\ov x_j-x_k}) +O(\epsilon ), & &
\ll{wexpand}\eea
Denoting $w_\pm=\ha (w_1 \pm w_2)$, which  
reduce 
   the elliptic 
Gaudin Hamiltonian 
   to
 $
H_n=  {1 \ov  \epsilon^2} H^{(2)}_n+
 {1 \ov  \epsilon} H^{(1)}_n+O( \epsilon ^0)$ with 
$H^{(a)}_0=-\sum_jH^{(a)}_j,  a=1,2$.
Therefore,    $H^{(a)}_j $
give mutually commuting independent  conserved quantities: 
 $[H_j^{(a)},H^{(b)}_k]=0 $, with
\be  H_j^{(2)}=S^3_j, \ \   H_j^{(1)}= H_j^{(bS)}+ H_j ^{(SS)}, \ j=1, \ldots, N_a,
 \ll{hjc}\ee
where
\be H_j^{(bS)}=
  \Omega \left( (bS^+_j+b^\dag
 S^-_j)+ (b^\dag S^+_j+b S^-_j)\right), \quad 
 \Omega ={ k^{'} \ov 2 \sqrt 2}, 
\ll{crwbs}\ee
 with  elliptic modulus $k= \sqrt {1- k^{'2}},$ describes 
matter-radiation interaction having explicit CRW terms, where    
by an allowed phase transformation $e^{i\pi \ov 2}$
 in both $b, S^+_j$,  we have
changed the sign of the RW term.   In (\re{hjc}) the part
\be H_j^{(SS)}=
\sum^{N_a}_{k \neq j} {1\ov x_j-x_k}( S^+_jS^-_k+ S^-_jS^+_k+ S^3_jS^3_k),
\ll{ess}\ee
accouts for interatomic interactions with inhomogenious coupling constants
expressed through $x_j$. These arbitrary parameters
may be adjusted to make  the interaction strengths diminishing with 
increasing distance between the atoms. 
Various  combinations of   Hamiltonians (\re{hjc}) with
 arbitrary coupling constants: $H_j=\omega_{j}S^3_j,+
J_j (H_j^{(bS)}+ H_j ^{(SS)} )$
can yield different   {\it integrable multi-atom matter-radiation models 
 without RWA}. For example, by taking  simply 
 $\sum_jH_j $ with $\omega_{j}=\omega , J_{j} =1$,
 we get an {\it integrable multi-atom JC-type  model 
with nontrivial CRW terms}:
\be H_0= \sum_j^{N_a}\left( \omega S^3_j + \Omega 
(b+b^\dag) ( S^+_j+ 
S^-_j) \right)\ll{crw0}\ee
which in contrast to the popular  belief \c{rwa} 
is exactly solvable through Bethe ansatz, as we see below.
To make the model more physical we should 
 add a field term $\omega_f b^\dag b $ to the Hamiltonian (\re{crw0}).
Considering the field frequency  $\omega_f$ to be small
we can treat this additional term  perturbatively 
 over the Bethe solvable original integrable
model. It is curious to  note that, essentially 
at this  small   field frequency limit   the 
RWA becomes a bad approximation and   the
CRW terms, which  we have considered,   become  significant.

\section{Exact Bethe ansatz solutions}

In the basic algebraic Bethe ansatz (ABA) method the trace of the 
 monodromy matrix  
 $\tau (\la)=tr T(\la)$ produces  conserved operators, while the
 off-diagonal elements $  T_{21}(\la)\equiv B(\la)$ and $ 
  T_{12}(\la) \equiv C(\la)$ 
   act like
   { creation} and { annihilation} operators of 
 pseudoparticles,  with the M-particle  state given as 
 $|M>_B=B(\la_1) \cdots B(\la_M)|0>$, with  the pseudovacuum 
   $|0>$  defined  through
$ C(\la)|0>=0$.
The major aim of ABA \c{aba} 
is to find the eigenvalue solution:
$\tau(\la)|M>_B=\Lambda (\la, \{ \la_a\})|M>_B$.

All our models with RWA, constructed in a unified way 
through the ancestor model scheme,  can   be  solved  
 in a unified manner
 following the ABA method in its standard formulation \c{aba}.
Omitting  here the details \c{jpa04}, we focus 
on the  ABA solution    of our integrable   models without
RWA 
(\re {hjc}- \re {crw0}). For deriving the ABA relations of these models
we have  to repeat  the steps
we have followed  in obtaining their Hamiltonians. That is 
we have   to start from the 
ABA relations for the $xyz$ spin chain with arbitrary spins ,
where a gauge transformed version of operators and states has to be
used to define a modified pseudovacuum \c{xyz,takabe}. 
Considering carefully the limit $\al \to 0$  the corresponding
ABA relations for the elliptic Gaudin model is obtained \c{egaudin}.
 Taking  the high spin limit for the radiative mode 
in these relations we derive finally  the eigenvalues and the Bethe
equations
 for our integrable models
without RWA.
  
 We start therefore 
from the 
 ellipic Gaudin model, and using  the key ABA relations  
from \c{egaudin}, we derive
the energy spectrum as
\be E^{(egaud)}_n= 2s_n \th _{11}^{'}(0) (i \pi \nu + \sum_{b=1}^M
 { 
\th _{11}^{'}(z_n-\la_b) 
\ov  \th _{11}(z_n-\la_b) }+\sum_{m \neq n}^Ms_m{ 
\th _{11}^{'}(z_n-z_m) 
\ov  \th _{11}(z_n-z_m) })
\ll{evgaudin}\ee
with   arbitrary spin values  $s_n$ of operators at the $n$-th
site and involving elliptic function  $\th _{11} $. 
The corresponding Bethe
equations
can be found in \c{egaudin} as
\be 
 \sum_{n=1}^N s_n
 { 
\th _{11}^{'}(\la_a-z_n) 
\ov  \th _{11}(\la_a-z_n) }=-i \pi \nu+
 \sum_{b \neq a}^M 
 { 
\th _{11}^{'}(\la_a-\la_b) 
\ov  \th _{11}(\la_a-\la_b) }
\ll{begaudin}\ee
Choosing the inhomogeneity parameters at $n=0,j$ as in (\re {z2x})
, introducing a scaling also for the
   Bethe parameters: $\la_a =\epsilon l_a$ and using some properties of the 
$\th _{11}$ function,  we can derive now  
 from  (\re{evgaudin}) for $n=j$ the exact eigenvalue  
for our matter-radiation model (\re{hjc}) without RWA,
at the high spin limit  (or equivalently, at $\epsilon \to 0$)
for the radiation field as
\be E^{(crw)}_j= 2s_j \th _{11}^{'} \left(\sum_{b=1}^M { 1 \ov  (x_j-l_b) }
+\sum_{k \neq j}^{N_a} { s_k \ov  (x_j-x_k)}+{\th _{10}^{''}\ov \th _{10}} 
(x_j-x_0) \right)
\ll{evcrw}\ee
where we have used short-hand notations $\th^{(l)} _{11}(0)=\th^{(l)} _{11}$,
for $ l=0,1,2. $
 Similarly expanding  (\re{begaudin})  
 we extract the corresponding  Bethe equation
\be 
{\th _{10}^{''}
\ov  \th _{10} }(x_0-l_a)+
 \sum_{k=1}^{N_a} s_k
 {1  
\ov  (x_k-l_a) }=
 \sum_{b \neq a}^M 
 { 1 
\ov  (l_b-l_a) }, \quad a=1,\ldots,M,
\ll{becrw}\ee
the solutions of which should  determine  the Bethe momentums 
 $l_b, b=1,\ldots,M$
for a given arbitrary nondegenerate
 set of inhomogeneity parameters  $x_j, j=1,\ldots,N_a$.
 Note that $x_0$
can be absorbed by shifting the parameters $x_j$ and $l_a$.

Either expanding  (\re {evgaudin}) for $n=0$ and 
combining terms of different orders,
or by taking the sum of  
 (\re {evcrw}) with an additional term with $\omega$
 and using the Bethe equation 
(\re {becrw}),  we can obtain  the 
exact eigenvalues
 for our multi-atom model (\re {crw0}), with explicit 
CRW terms as
\be E^{(crw)}_0= { \th _{11}^{'}\th _{10}^{''}
\ov  \th _{10} } \left(\sum_{b=1}^M {  (x_0-l_b) }
+\sum_{k =1}^{N_a}  s_k  (x_0-x_k)\right)
+2i \omega  \pi\th _{11}^{'} \nu
\ll{evcrw0}\ee

 \section{Concluding remarks}

In real matter-radiation models involving many atoms interatomic
interactions must appear. Like-wise 
the customary rotating wave approximation  in general must be avoided, at
least  when 
 away from the resonance point.
However these important cases   could not be included without spoiling the
integrability  
in most of the solvable models proposed earlier. 
We have proposed new integrable models, where  
 both these cases  
 can be incorporated retaining the   integrability of the system and extract
exact   solutions through    
   the Bethe ansatz.
For models with rotating wave approximation we  find   
 a series of  integrable multi-atom models with 
 inter-atomic interactions, which gives generalizations of 
   Jaynes-Cummings, Buck-Sukumar and trapped ion models
together with their q-deformations.
These models belong to the rational or the trigonometric class 
and can be solved exactly in a unified way.

  We also construct for the first time  exactly  solvable 
 multi-atom  matter-radiation models without rotating wave approximation
(RWA), which are
 close to the physical systems with
 the required   counter rotating wave (CRW) terms.
 We achieve this  by taking  the high spin limit for the radiation field
in  the elliptic
Gaudin model.
However in contrast to
our models  with RWA, where we could construct 
different types of matter-radiation models,
our success with models without RWA is  restricted, where  we could
obtain 
 only  JC type  models  with CRW terms. Moreover,
  the free radiation term
 does not  naturally  appear in such integrable models and has to
 be included
 perturbatively over the exact BA result.

Identifying the models in real systems
and  experimental verification  of
  the related results presented here, especially in many-atom microlasers
\c{n03} or in trapped ions away from  resonance point \c{rwa}
 would be an important application.

 \end{document}